\documentclass[prl
,twocolumn
,floatfix
amsmath,amssymb,showpacs
]
{revtex4}

\usepackage{graphicx,color,amsmath}
\usepackage{array}
\usepackage{subfigure}
\graphicspath{{Diagrams/}}

\newcommand{\beq}{\begin{equation}}
\newcommand{\eeq}{\end{equation}}
\newcommand{\bqa}{\begin{eqnarray}}
\newcommand{\eqa}{\end{eqnarray}}

\newcommand{\erf}[1]{Eq.~(\ref{#1})}

\newcommand{\dg}{^\dagger}

\newcommand{\cu}[1]{\left\{ {#1} \right\}}
\newcommand{\ro}[1]{\left( {#1} \right)}

\newcommand{\ket}[1]{\ensuremath{\left| #1 \right\rangle}}
\newcommand{\bra}[1]{\ensuremath{\left\langle #1 \right|}}
\newcommand{\braket}[2]{\ensuremath{\left\langle #1|#2 \right\rangle}}

\newcommand{\ea}{{\em et al.}}

\definecolor{nblue}{rgb}{0.2,0.2,0.7}
\definecolor{ngreen}{rgb}{0.2,0.6,0.2}
\definecolor{nred}{rgb}{0.7,0.2,0.2}
\definecolor{nblack}{rgb}{0,0,0}

\renewcommand{\section}[1]{{\em #1}. --- }

\begin{document}

\title{How many bits does it take to track an open quantum system?}
\author{R. I. Karasik} \email{R.Karasik@Griffith.edu.au}
\author{H. M. Wiseman}\email{H.Wiseman@Griffith.edu.au}
\affiliation{Centre for Quantum Computer Technology, Centre for Quantum Dynamics, Griffith University, Brisbane, Queensland 4111, Australia}

\begin{abstract}
A $D$-dimensional Markovian open quantum system will undergo  quantum jumps between pure states, if we can monitor the bath to which it is coupled with sufficient precision.
In general these jumps, plus the between-jump evolution, create a trajectory which passes through infinitely many different pure states.  Here we show that, for any ergodic master equation, one can expect to find an {\em adaptive} monitoring scheme on the bath that can confine the system state to jumping between only $K$ states, for   some  $K \geq (D-1)^2+1$. For $D=2$ we explicitly construct a 2-state ensemble for any ergodic master equation, 
showing that one bit is always sufficient to track a qubit.
\end{abstract}


\maketitle

The first quantitative model of quantum {\em dynamics} was 
Einstein's model of stimulated and spontaneous {\em jumps} \cite{Ein17} between 
Bohr's stationary atomic states \cite{Boh13}. In modern language, this is a model for 
an open quantum system weakly coupled to a heat bath, and, as Einstein showed, 
such jumps can lead to an equilibrium state that is a thermal mixture of energy eigenstates, 
with finite entropy. In Einstein's model, if one could track the individual stochastic events of 
energy exchange between atom and bath, then one would know which energy eigenstate 
the system occupied at any time. Truncating to a finite number $D$ of 
energy eigenstates, it follows that only a finite classical memory 
is required to keep track of the quantum system (that is, to know its exact pure state) 
in thermal equilibrium: a $K$-state memory with $K = D$.

Einstein's theory is a special case of Markovian 
open quantum system dynamics for finite-dimensional systems, which most generally are 
describable by a Lindblad-form master equation (ME)  \cite{WisMil10}: 
\beq \label{me1}
\dot{\rho} = {\cal L}\rho \equiv -i[  \hat H_{\rm eff}\rho   -  \rho\hat H_{\rm eff}\dg] 
+ \sum_{l=1}^L \hat c_l \rho \hat c\dg_l,
\eeq
where $ \hat H_{\rm eff} \equiv \hat H - i\sum_l  \hat c\dg_l \hat c_l/2$.  
Here $\hat H$ is Hermitian (it is the Hamiltonian) 
but the  jump operators $\{\hat c_l\}$ are completely arbitrary. 
Einstein's theory is a special case because in it 
each jump operator is proportional to $\ket{E}\bra{E'}$, for 
some $\hat H$-eigenstates $\ket{E}$ and $\ket{E'}$, so that the state after any jump is a stationary 
state $\ket{E}$. For a general ME, it is always possible, in principle, to monitor 
the bath such that every jump is resolvable, so that the system can be known to be in some pure state $\ket{\psi(t)}$ at all times \cite{Car93b,Wis96a,WisMil10}. 
However, in general, after a jump at time $\tau_j$, the state $\propto \hat c_l\ket{\psi(\tau_j)}$  will depend on the pre-jump state $\ket{\psi(\tau_j)}$, and will not be an an eigenstate of $\hat H$. Even if it were an energy eigenstate, it would not 
 in general  remain stationary until the next jump, because its subsequent 
evolution would be generated by the effective (non-Hermitian) Hamiltonian $\hat H_{\rm eff}$ appearing in \erf{me1}. 

It is thus not at all obvious whether for a general finite-dimensional open quantum system it would be possible to keep track of its pure state, even in principle,  with a finite classical memory. On the face of it, it would seem necessary to store the 
exact times of each jump --- a  sequence of real numbers $\{\tau_j:j\}$ each of which would require, in principle, an infinite memory to store. Alternately one could store the conditioned quantum state   $\ket{\psi(t)}$ itself, but this (a $D$-dimensional complex vector) would also require an infinite memory. This situation is of course completely different from a finite-state stochastic classical system  (which is what Einstein's model amounts to),  where a finite-state classical memory of the same size is always sufficient. 

In this Letter we address this fundamental question about open quantum system dynamics. We show that for any 
ergodic Markovian dynamics \footnote{Here ergodic simply means that the master equation has a unique stationary state.} 
of a  $D$-dimensional quantum system, one can expect to be able to track the state with a $K$-state classical apparatus   for some $K  \geq  (D-1)^2+1$. This is possible only because there is {\em entanglement} between the system and bath, which means that different monitoring schemes on the bath  give rise to different sorts of stochastic pure state trajectories (``unravellings'' \cite{Car93b}) for a given ME.  We then prove that for $D=2$ (a qubit), $K=2$ is always sufficient; that is, there is always an unravelling for which the qubit jumps between only two possible states, $\ket{\phi_1}$ and $\ket{\phi_2}$. Although this sounds similar to Einstein's dynamics, it is in fact quite different in general --- the two states are {\em non-orthogonal}, $\braket{\phi_1}{\phi_2} \neq 0$, and the monitoring of the qubit's environment must be {\em adaptive}, controlled by the classical bit that stores the state of the qubit. 

We begin by revisiting {\em the preferred ensemble fact} \cite{WisVac01}, to explain why it is not possible in general to unravel a  ME  such that the system jumps between the eigenstates of the equilibrium $\rho$ (as in Einstein's model). Then we  show  the general result cited above for $D$-dimensional systems, and give an explicit construction of the adaptive unravelling for the special case of cyclic jumps with only one Lindblad operator $\hat c$. We then prove that one bit is always sufficient to track a qubit,  and that in  some cases it is actually possible to store the state of the open qubit using {\em less than one bit} of memory {\em on average}, if we imagine an ensemble of $N\gg 1$ qubits, each independently measured.  Surprisingly, considering $K>2$ can actually help in this regard.  We illustrate these phenomena using the resonance fluorescence  ME \cite{Car93b}.

\section{The preferred ensemble fact} 
Consider  a  Linbladian $\cal{L}$ with unique steady state defined by ${\cal L}\rho_{\rm ss}=0$ and we assume that $\rho_{\rm ss}$ is a mixed state. 
This mixed state can be decomposed in terms of pure states $\ket{\phi_k}$ via $\rho_{\rm ss}= \sum_{k=1}^K  \wp_k \ket{\phi_k}\bra{\phi_k}$  with positive constants $\wp_k$. Note that there are infinitely many such decompositions,  as the  states $\ket{\phi_k}$ need not be orthogonal. However, only for some decompositions is it  possible to devise a way to monitor the  system's environment --- which leaves the average evolution of the system unchanged from Eq.~(\ref{me1}) --- such that  the system will only ever be in one of the states $\ket{\phi_k}$, and will spend a proportion of time in that state equal to $\wp_k$ in the long-time limit. 
Decompositions  $\{\wp_k , \ket{\phi_k}\}$  that can be realized in this way  are called {\em physically realizable} (PR).  The fact that some decompositions are not PR is known as {\em the preferred ensemble fact} \cite{WisVac01}.

As shown in Ref.~\cite{WisVac01}, an ensemble $\{ \wp_k,\, \ket{\phi_k} \}$ is PR iff (if and only if) there exists rates $\kappa_{jk} \geq  0$ such that 
\beq
\label{jumpCond} 
  \forall k, \ {\cal L}\ket{\phi_k}\bra{\phi_k} = \sum_{k=1}^K \kappa_{jk} 
 \left(\ket{\phi_j} \bra{\phi_j}-\ket{\phi_k} \bra{\phi_k} \right).
\eeq
For a general  ME,  most  decompositions $\{ \wp_k,\, \ket{\phi_k} \}$  of $\rho_{\rm ss}$ are not PR, including the $K=D$ ensemble composed from the diagonal basis for $\rho_{\rm ss}$ \cite{WisMil10}.

\section{The existence of PR ensembles}
For finite $K$ and $D$, searching for solutions of \erf{jumpCond} reduces to solving polynomial equations. We can describe $K$ pure states with $K(2 D-1)$ real unknowns  and $K$ quadratic constraints from normalization.  Eq.~(\ref{jumpCond}) introduces $K^2-K$ unknown rates $\kappa_{jk}$ and imposes an additional $K (D^2-1)$ cubic constraints (the minus one is because both sides are traceless by construction).  Thus we have $K D^2$ polynomial constraints  and $K(2D+K-2)$ unknowns. For $K>(D-1)^2+1$ we have an underdetermined system of equations. For linear underdetermined systems, given by equations $\{f_j\}$,  there are infinitely many solutions  except for the set of measure zero for which there exist constants $\{\alpha_j\}$  such that $\sum_j \alpha_j f_j+1=0$. Similarly, for polynomial systems $\{p_j\}$, the Real Nullstellensatz~\cite{Sturm} certifies that there are no real solutions iff there exist some polynomials $\{a_j\}$ and $\{d_k\}$ such that $\sum_j a_j p_j+\sum_k d_k^2+1=0$.  Thus, for a general ME we expect to be able to find a $K$-element PR ensemble for some  $K-1  \geq   (D-1)^2$.

The freedom that experimentalists have (in principle) to realize different PR ensembles comes from the ability to 
monitor the system's environment in different ways. This can be understood as follows. The ME  (\ref{me1}) is invariant under the transformations $\cu{\hat c_l} \to \cu{\hat c'_m}$,  where  $\hat c'_m  = \sum_{l=1}^{L} S_{ml}\hat c_l + \beta_m$, and $\hat H \to \hat H'= \hat H - \frac{i}{2}\sum_{m=1}^{M}  (\beta^*_m  \hat c'_m - \beta_m \hat c'_m{}\dg)$. Here $\vec\beta$ is an arbitrary complex vector and 
${\bf S}$ is an arbitrary semi-unitary  matrix --- $\sum_{m=1}^{M} S_{l'm}^* S_{ml} = \delta_{l',l}$.  	
Unravelling this ME with $\cu{\hat c'_m}$ as the jump operators and $\hat H'_{\rm eff} = \hat H' - i \sum_{m=1}^M \hat c'_m{} \dg \hat c'_m /2$ as the effective non-Hermitian Hamiltonian clearly gives different stochastic evolution, while leaving the average evolution unchanged. 
To obtain the most general pure-state unravelling of the  ME,   we require $\vec\beta$ and ${\bf S}$ to depend upon the previous record of jumps. That is, we require an adaptive monitoring \cite{Wis96a,WisMil10}. Of course when we use this to achieve jumping between a finite number of states, the classical $K$-state memory that stores which state the system is currently in carries all the information necessary for determining $\vec\beta$ and ${\bf S}$. That is, the adaptive unravelling is specified by  $K$  different values for $\vec{\beta}$ and ${\bf S}$. The physical meaning of these parameters is most easily explained in a quantum optics context: ${\bf S}$ describes a linear interferometer \footnote{Here `linear interferometer' is to be understood in the most general sense, including frequency shifters if the system has outputs in different frequency bands.} taking the field outputs from the system as inputs, while $\vec\beta$ describes adding (weak) local oscillators to the output fields from the interferometer prior to detection by photon counting. Recently, adaptive control of a weak local oscillator has been used 
for optimally distinguishing coherent states \cite{JM07}, and of a strong local oscillator for improved phase estimation \cite{Whe10}.

\section{Backing out the measurement scheme} 
Although for every PR ensemble there must exist a monitoring scheme exists by definition, it may not be easy to find.  We now present an explicit method for determining this scheme for the special case of cyclic jumps with a single Lindblad operator $\hat c$. That is,  we assume that the system in the state $\ket{\phi_{k}}$ always jumps  to the state $\ket{\phi_{k+1}}$ (strictly, $\ket{\phi_{(k+1)\,\text{mod}\,K}}$). In this case there are only $K$ jump rates, so the number of real unknowns is only $2KD$.  With a single Linblad operator and cyclic jumps, both sides of \erf{jumpCond} have rank two by construction, so for $D>2$ \erf{jumpCond} is less constraining than in the general case. Nevertheless, the system will be overconstrained for $D>2$. Thus in general we do not expect there to exist cyclic jump solutions for $D>2$. Later we exhibit eleven different cyclic jump solutions for a qubit ($D=2$), and the method here is applicable to each of them. 

In the case of a single Lindblad operator, the only freedom in the unravelling is in choosing $\beta^k$, the local oscillator amplitude when the system is known to be in state $\ket{\phi_k}$  (that is, when the $K$-state classical memory is in state $k$).
This gives the jump operator $\hat c +  \beta^k$ and the effective Hamiltonian
$\hat H_{\rm eff}^k = \hat H_{\rm eff} + i\beta^k{}^* \hat c - i|\beta^k|^2 /2$. 
Thus the system will undergo cyclic jumps iff
\beq  \label{cycliciff}
(\hat H_{\rm eff} + i\beta^k{}^* \hat c) \ket{\phi_k} \propto \ket{\phi_k}\;,\;\; (\hat c + \beta^k) \ket{\phi_k} \propto \ket{\phi_{k+1}},
\eeq
Now by assumption  from \erf{jumpCond},  ${\cal L}\ket{\phi_k}\bra{\phi_k} \propto \ket{\phi_{k+1}} \bra{\phi_{k+1}}-\ket{\phi_k} \bra{\phi_k}$. From this we can show that 
\begin{align}
\hat{c} \ket{\phi_k}& = a_k\ket{\phi_k} + b_k\ket{\phi_{k+1}} , \\
\hat{H}_{\rm eff}\ket{\phi_k}& = c_k\ket{\phi_k} +  i a_k^*b_k\ket{\phi_{k+1}}.
\end{align}
For some coefficients $a_k$, $b_k$ and $c_k$ that are easily found given $\ket{\phi_k}$ and $\ket{\phi_{k+1}}$. Comparing this to \erf{cycliciff}, we see that   choosing $\beta^k = - a_k$ gives cyclic jumps as required.

\section{For a qubit, one bit is all it takes} 
We now prove that a 2-state PR ensemble always exists for a qubit. 
We use the Bloch representation, so that \erf{me1} becomes
\beq
\dot{\vec{r}}=A\vec{r}+\vec{b}, \label{bve}
\eeq 
where  $A$ is a $3\times3$-matrix and $\vec{b}$ is a 3-vector. As always, we assume that there exists a unique steady state $\vec{r}_{\rm ss}=-A^{-1}\vec{b}$, which is the case   iff the real part of each eigenvalue of $A$ is  negative. We can track this system with a $K$-state memory iff there exists  an ensemble $\{\wp_k,\,\vec{r}_k\}$ and rates $\kappa_{jk}\geq 0$ such that 
\begin{align}
\forall k,\ \vec{r}_k\cdot\vec{r}_k=&1, \label{bvNormCond}\\
\forall j, \ A\vec{r}_j+\vec{b}=&\sum_{k=1}^K \kappa_{jk}(\vec{r}_k-\vec{r}_j).\label{bvJumpCond}
\end{align}
Thus the problem reduces to finding a real solution to a system of quadratic  equations with real coefficients. 
This type of problem is surprisingly hard even for a small number of unknowns, and is known to be an NP-complete problem in general \cite{BCSS}.

Luckily, the simplest case of $K=2$ has an analytical solution. Here the 
qubit is assumed to jump between two states, $\vec{r}_1$ and $\vec{r}_2$. Then Eqs.~(\ref{bvJumpCond}) reduce  to  a single equation, $A(\vec{r}_1-\vec{r}_2)=(\kappa_{12}+\kappa_{21})(\vec{r}_2-\vec{r}_1)$,  which is simply an eigenvalue equation. Thus $\vec{r}_1-\vec{r}_2=\vec{v}$ is an eigenvector of $A$ with eigenvalue $\lambda = -(\kappa_{12}+\kappa_{21})$. From \erf{bvNormCond} and the fact that $\wp_1 = {\kappa_{21}}/\ro{\kappa_{21}+\kappa_{12}} = 1-\wp_2$, it is simple to show that $\vec{r}_1$ and $\vec{r}_2$ have the form
\begin{align}
\vec{r}_1=&\vec{r}_{\rm ss}+\check{v}\sqrt{1-\|\vec{r}_{\rm  ss}\|^2}\sqrt{{\wp_1}/\ro{1-\wp_1}}\label{r1}\\
\vec{r}_2=&\vec{r}_{\rm  ss}-\check{v}\sqrt{1-\|\vec{r}_{\rm  ss}\|^2}\sqrt{\ro{1-\wp_1}/{\wp_1}}\label{r2}
\end{align}
with $\check{v}=\vec{v}/\|\vec{v}\|$ is the normalized eigenvector, and 
\beq
\wp_1=\frac{1}{2}\left(1-\frac{\langle \vec{r}_{\rm  ss}, \check{v}\rangle}{\sqrt{1-\|\vec{r}_{\rm  ss}\|^2+\langle \vec{r}_{\rm  ss}, \check{v}\rangle^2}}\right) = \frac{\kappa_{21}}{|\lambda|}.
\eeq
Because the Bloch vectors must be real, only real eigenvectors $\vec{v}$ of $A$ can contribute to the solution.  
By assumption, $A$ has three nonzero eigenvalues and, by  the fundamental theorem of algebra, at least one eigenvalue (and consequently one eigenvector) is real. Therefore, a qubit always has a preferred ensemble comprising just two states.

\section{Entropy} As  nooted above, it may be possible to store the state of a qubit in {\em less than} one bit, in an average sense. We can quantify this by using the Shannon entropy. Under continuous monitoring, in the long-time limit,  the system will occupy states $\ket{\phi_k}$ with probabilities $\wp_k$. The Shannon entropy for  this ensemble  is  $h\left(\{\wp_k\}\right)=-\sum_k \wp_k\log_2 \wp_k$.  This is lower bounded by the von Neumann entropy for the  steady-state mixture: 
\beq \label{boundonh}
h\left(\{\wp_k\}\right) \geq S(\rho_{\rm ss}) \equiv -\text{Tr}[\rho_{\rm  ss}\log_2 \rho_{\rm  ss}],
\eeq
with equality iff $\{\wp_k, \ket{\phi_k}\bra{\phi_k}\}$ is the diagonal ensemble. 
 
Note that if the eigenvector $\vec{v}$ of matrix $A$ used to construct  the  ensemble in Eqs.~(\ref{r1}-\ref{r2}) is orthogonal to the steady state $\vec{r}_{\rm ss}$, then the probability of occupying states $\vec{r}_1$ and $\vec{r}_2$ is $1/2$. In this case  the Shannon entropy is 1, meaning that one bit is sufficient to track the state of the system.  If, on the other hand, $\vec{v}$ is not orthogonal to $\vec{r}_{\rm ss}$, then the  Shannon entropy $h$ for the ensemble will be less than one and one could store the state on the qubit in \emph{less than} one bit \emph{on average}. That is, one could keep track of the state of a collection of $N$ identically monitored qubits using only $N h$ bits, in the limit of large $N$. 

\section{Resonance fluorescence} 
\begin{figure} [t]
\includegraphics[width=.95\columnwidth]{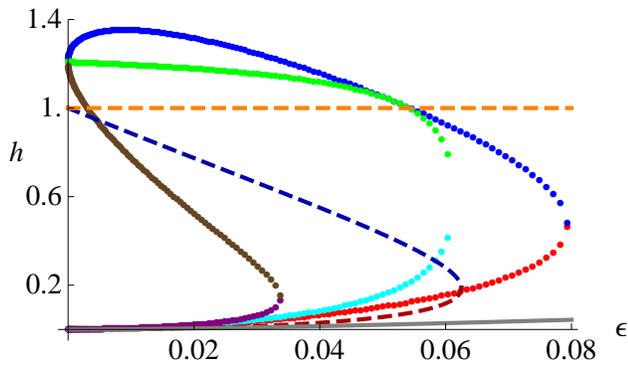} 
\vspace{-2mm}
\caption{\label{graph} 
The  average number of bits required   to  keep track of the pure state of  a qubit 
described by \erf{Ab}, as a function of dimensionless driving power $\epsilon=\Omega^2/\gamma^2$. 
The solid line is the von Neumann entropy for $\rho_{\rm ss}$, a lower bound on the memory required. The other nine curves are for  eleven  different adaptive unravellings. Dashed lines are for 2-state jumping. Dotted lines are for cyclic 3-state jumping. 
 }
\vspace{-4mm}
\end{figure}

In order to illustrate our ideas, we consider  the example of resonance fluoresence of a two-level atom (a qubit)  with basis states $\ket{0}$ and $\ket{1}$. The atom is coupled to the continuum of electromagnetic radiation and so decays to $\ket{0}$ at rate $\gamma$. At the same time, it is driven by a classical field with Rabi frequency $\Omega$. The  qubit evolution in the interaction frame is given by a ME of the form of Eq.~(\ref{me1}) with $\hat H=\Omega(\ket{0}\bra{1}+\ket{1}\bra{0})/2 $ and one jump operator $\hat c = \sqrt{\gamma}\ket{0}\bra{1}$ \cite{WisMil10}.
For this case, $A$ and $\vec{b}$ in the Bloch vector equation, Eq.~(\ref{bve}), are
\begin{equation} \label{Ab}
 A=\begin{pmatrix} 
    -\gamma/2& 0&0\\
    0&-\gamma/2&-\Omega\\
    0&\Omega&-\gamma
   \end{pmatrix}\,\,\, \text{ and }
\vec{b}=\begin{pmatrix}
         0\\ 0\\ \gamma
        \end{pmatrix}. 
\end{equation}
The steady state $\vec{r}_{\rm ss}= (0, 2\gamma\Omega,-\gamma^2)^T/\ro{\gamma^2+2\Omega^2}$, is a mixed state for $\Omega\neq0$. 
The eigenvectors of $A$ are $\vec{v}_1=(1,\,0\,0)^T$ and $\vec{v}_\pm=(0, \gamma\pm\sqrt{\gamma^2-16 \Omega^2}, 4 \Omega)^T$. Eigenvector $\vec{v}_1$ is real and orthogonal to  $\vec{r}_{\rm ss}$.  Thus it will always yield a solution with Shannon entropy $h=1$. This solution was originally discovered in  Ref.~\cite{WT99}. Eigenvectors $\vec{v}_\pm$ are real only for $|\Omega|<\gamma/4$ and yield solutions with $h<1$. In fact,  for $\epsilon\equiv \Omega^2/\gamma^2 \ll 1$,  the entropy of the solution due to $\vec{v}_-$  differs from $S(\rho_{\rm ss})$ only at $O(\epsilon^3)$.  
Thus for small driving, this ensemble has an entropy very close to the bound (\ref{boundonh}). This is seen in Fig.~\ref{graph}, where dashed lines show $h$ 
for the three different 2-state solutions. 

For $\epsilon>0.0625$ (that is, $|\Omega|>\gamma/4$) there are no low-entropy 2-state solutions. Surprisingly, by increasing  $K$, the number of states,  from 2 to 3, we regain a relatively low entropy solution for some range of $\epsilon>0.0625$. 
Recall that for 2-state jumping, we use one {\em real} eigenvector of $A$ to construct the PR ensemble, as the  Bloch vectors must be co-linear with $\vec{r}_{\rm ss}$.  For 3-state jumping,  the Bloch vectors must be coplanar (and {\em not} colinear) with  $\vec{r}_{\rm ss}$,  so we require two eigenvectors.  When  $A$ has complex eigenvectors, they come in conjugate pairs, and  we can, for some values of $\Omega$, construct a PR ensembles with 3 real Bloch vectors using these conjugate pairs.

We find all possible 3-state cycles by numerical search for solutions to Eqs.~(\ref{bvNormCond}-\ref{bvJumpCond})   using symbolic-numerical algorithms based on computing Groebner bases \cite{CLO}. 
There are eight solutions, coming in pairs, as shown (as dotted lines) in Fig.~\ref{graph}. 
For $\epsilon<0.0795$, there are two PR ensembles generated from complex eigenvectors $\vec{v}_\pm$ of $A$.  
These give the highest and lowest of the 3-state jumping entropy curves. In the region  $\epsilon<0.0610$, there are an additional four solutions constructed from $\vec{v}_1$ and $\vec{v}_-$. Only two new curves appear in Fig.~\ref{graph} because they come in degenerate pairs. Finally, for $\epsilon<0.0335$, there are two more solutions constructed from $\vec{v}_\pm$. 
As $|\Omega|$ decreases, the entropy for four of the solutions approaches 1.206,  whereas the entropy for the other four approaches 0.

\begin{figure}
\subfigure{
 \includegraphics[width=4.125cm, height=4.125cm, trim=10 10 10 10]{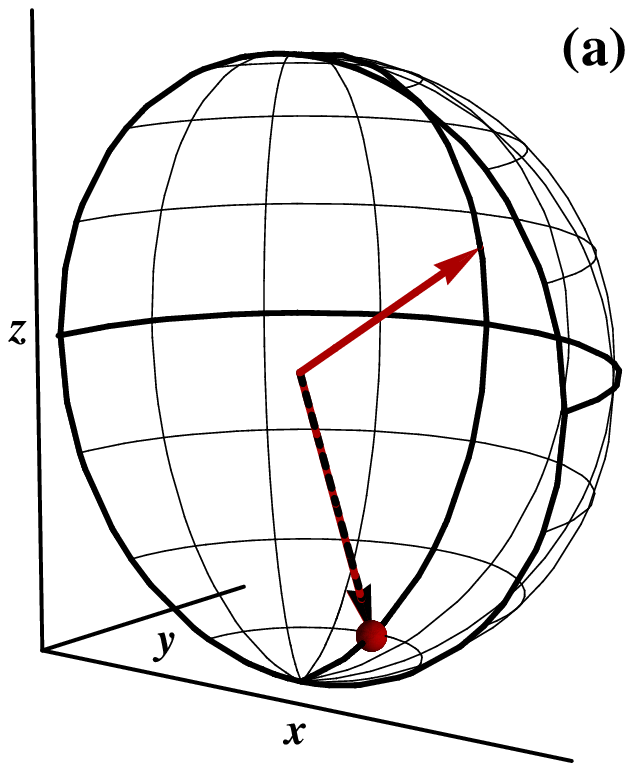}
}\hspace{-2.5 mm}%
\subfigure{
 \includegraphics[width=4.125cm, height=4.125cm, trim=10 10 10 10]{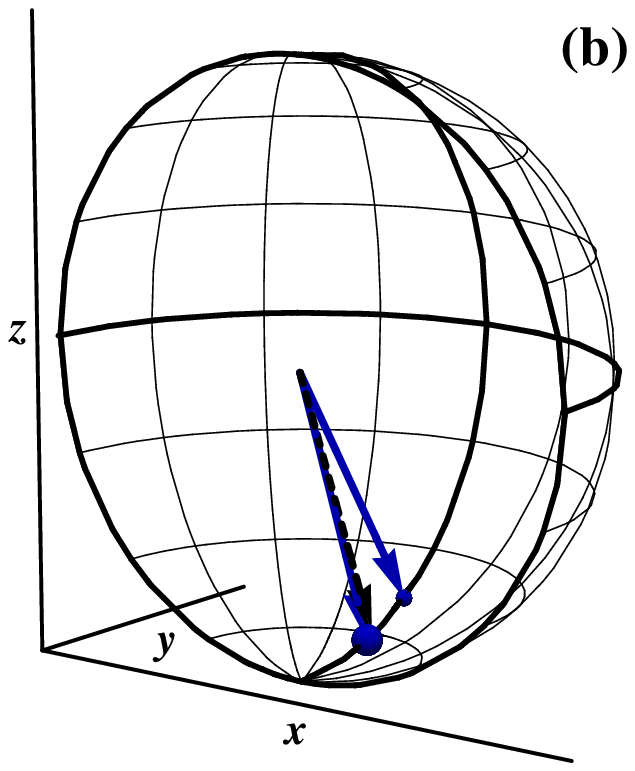}
}\vspace{-5 mm}
\subfigure{
 \includegraphics[width=4.125cm, height=4.125cm, trim=10 10 10 10]{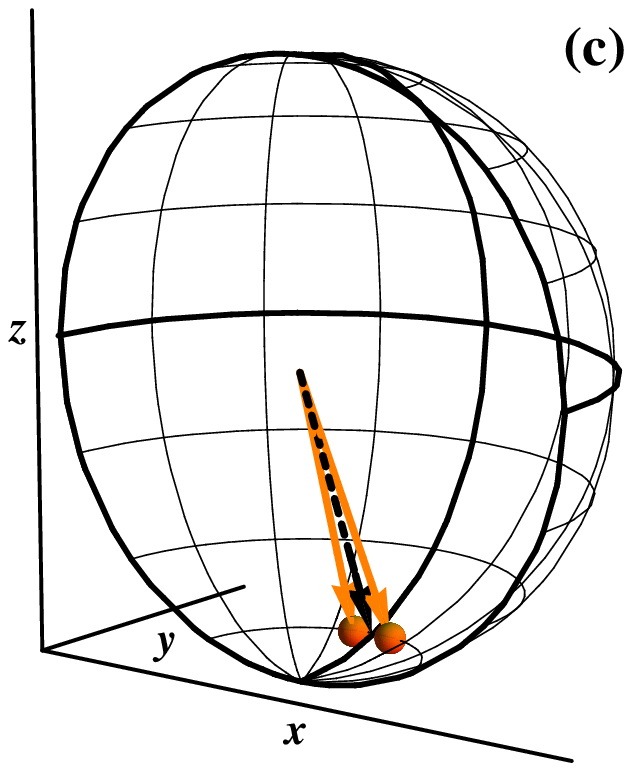}
}\hspace{-2.5 mm}%
\subfigure{
 \includegraphics[width=4.125cm, height=4.125cm, trim=10 10 10 10]{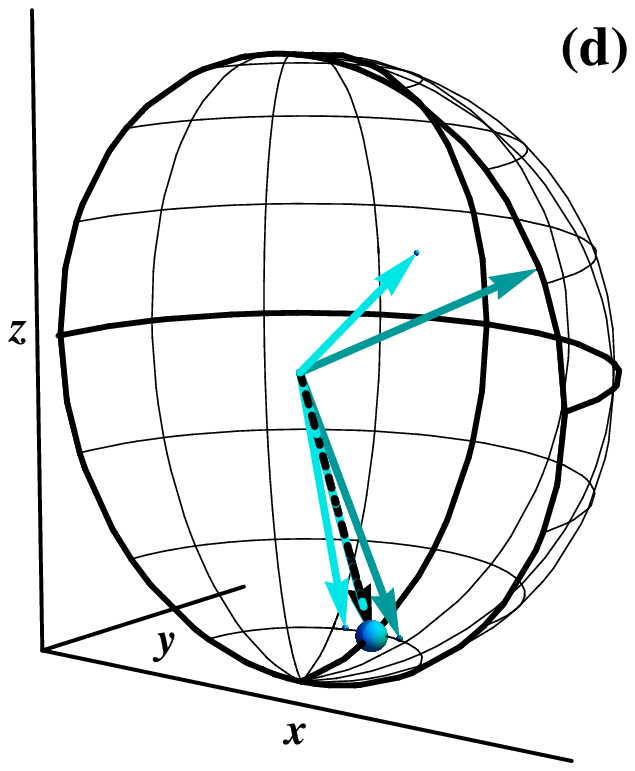}
}\vspace{-5mm}
\caption{Solid arrows show Bloch vectors for 2-state jumping (a)-(c), and 3-state jumping (d). Colors [online] match those of Fig.~\ref{graph}. The volume of the sphere at the tip of each arrow represents the probability that the qubit occupies the corresponding pure state.  The dashed arrow is $\vec{r}_{\rm ss}$. For all plots, $\Omega/\gamma=0.2$ ($\epsilon=0.04$). }
\label{bv}
\vspace{-5mm}
\end{figure}

The three 2-state PR ensembles and two of the 3-state PR ensembles are shown in Fig.~\ref{bv}. Figure~\ref{bv}(a) shows the main features of low-entropy solutions: the  states are far apart, and the qubit spends most of the time in one state that is nearly aligned with the steady state. Figure~\ref{bv}(b) captures the nature of high-entropy solutions: the states cluster around steady state. In both of these cases the ensemble lies in the $x=0$ plane. The ensemble with $h=1$, Fig.~\ref{bv}(c) does not, but is still symmetric under reflection in this plane. This symmetry is respected for all ensembles found except for the pairs of 3-state jumping solutions  that are degenerate  with respect to entropy.  Figure~\ref{bv}(d) shows one such pair: the two ensembles are mirror images of each-other in the $x=0$ plane.

In summary, we considered  an arbitrary ergodic Markovian open quantum system subject to continuous monitoring that resolves every jump and allows the system to stay in a pure state.  Under a generic monitoring scheme the system state will explore a manifold of pure states, so tracking it would require infinite memory. Here we showed that this situation is not an intrinsic property of open quantum systems, but is just a consequence of using the ``wrong'' monitoring scheme --- a finite ($K$-state) classical memory is sufficient to track the state of the system, by adaptively changing the scheme used to monitor the  environment, controlled by the state of the classical memory that stores the state of the quantum system. 
In general one would expected to need at least $K=(D-1)^2+1$ classical states to track a $D$-dimensional quantum system. 
The gap between $K$ and $D$ may be related to the recent result that there are stochastic processes that can be generated using quantum systems of lower dimensionality than is possible using only classical systems~\cite{MBW}.  The above quadratic difference, $K-1=(D-1)^2$, is also reminiscent of other comparisons between quantum and classical systems~\cite{Hardy}, so whether this $K$ is always sufficient is an important open question.  For $D=2$, however, the answer is now known, and is: yes, one bit is always enough to track the state of a qubit. 

This work was funded by the ARC grants CE0348250 and FF0458313.

\end{document}